\documentclass[onecolumn]{aastex63}

\usepackage{amsmath}
\received{December 11, 2019}
\revised{--}
\accepted{--} 
\submitjournal{PASP}

\shorttitle{Uncertainties of galaxy kinematics}
\shortauthors{Chilingarian \& Grishin}

\begin{document}

\title{Estimating Statistical Uncertainties of Internal Kinematics of Galaxies and Star Clusters Derived Using Full Spectrum Fitting\footnote{The source code is available in a public {\sc git} repository under the GPLv3 license: \url{https://bitbucket.org/extragalactic/pxf_kin_err}}}

\correspondingauthor{Igor Chilingarian}
\email{igor.chilingarian@cfa.harvard.edu, chil@sai.msu.ru}

\author[0000-0002-7924-3253]{Igor V. Chilingarian}
\affiliation{Smithsonian Astrophysical Observatory, 60 Garden St. MS09, Cambridge, MA 02138, USA}
\affiliation{Sternberg Astronomical Institute, M.V.Lomonosov Moscow State University, Universitetsky prospect 13, Moscow, 119234, Russia}

\author[0000-0003-3255-7340]{Kirill A. Grishin}
\affiliation{Sternberg Astronomical Institute, M.V.Lomonosov Moscow State University, Universitetsky prospect 13, Moscow, 119234, Russia}
\affiliation{Department of Physics, M.V. Lomonosov Moscow State University, 1 Vorobyovy Gory, Moscow, 119991, Russia}

\begin{abstract}
Pixel-space full spectrum fitting exploiting non-linear $\chi^2$ minimization became a \emph{de facto} standard way of deriving internal kinematics from absorption line spectra of galaxies and star clusters. However, reliable estimation of uncertainties for kinematic parameters remains a challenge and is usually addressed by running computationally expensive Monte-Carlo simulations. Here we derive simple formulae for the radial velocity and velocity dispersion uncertainties based solely on the shape of a template spectrum used in the fitting procedure and signal-to-noise information. Comparison with Monte-Carlo simulations provides perfect agreement for different templates, signal-to-noise ratios and velocity dispersion between 0.5 and 10 times of the instrumental spectral resolution. We provide {\sc IDL} and {\sc python} implementations of our approach. The main applications are: (i) exposure time calculators; (ii) design of observational programs and estimates on expected uncertainties for spectral surveys of galaxies and star clusters; (iii) a cheap and accurate substitute for Monte-Carlo simulations when running them for large samples of thousands of spectra is unfeasible or when uncertainties reported by a non-linear minimization algorithms are not considered reliable.
\end{abstract}


\section{Introduction and Motivation} \label{sec:intro}

Stellar and gas kinematics in galaxies keeps a fossil record of their evolution. By analyzing the motions of stars and comparing them to dynamical models, we can derive important properties of a stellar system such as its total mass, a density profile, a degree of rotational support \citep[see e.g.][]{Cappellari08}. By comparing observations against stellar population models, we can also get an insight on the stellar content of a galaxy or a star cluster. All this information can be extracted from spectra integrated along the line of sight. A pixel-space fitting approach, when a galaxy spectrum is approximated by a template and analyzed in every pixel along the wavelength axis was initially proposed by \citet{1992MNRAS.254..389R} and then slightly modified and greatly popularized by \citet{CE04}. Their ``penalized pixel-fitting'' or {\sc ppxf} technique based on a constrained non-linear $\chi^2$ minimization became a \emph{de-facto} standard way of extracting kinematics from absorption line spectra.

An important aspect of data analysis is the estimation of systematic and statistical uncertainties of parameters returned by a data analysis technique, because they can potentially render an obtained result statistically insignificant. Systematic errors, which might originate from an incomplete knowledge of how a dataset was obtained or from degeneracy between parameters of a model, can be very hard to assess. But even statistical uncertainties can be difficult to estimate, especially when a data analysis technique is complex. A typical approach is to use Monte-Carlo simulations to obtain statistical errors by directly analyzing a distribution of solutions from different noise realizations. Despite the simplicity of this approach, it is computationally intensive and the time required to properly sample a multi-dimensional parameter space skyrockets with the increasing number of dimensions.

Monte-Carlo simulations are recommended as a preferred method for estimating the uncertainties of galaxy kinematics in the {\sc ppxf} documentation, because the formal uncertainties returned by a constrained non-linear minimization procedure can sometimes be unreliable, e.g. when parameters are degenerated or some of them reach a limit set by a constraint. However, this can become prohibitively expensive when dealing with large samples of spectra. When we were preparing the Reference Catalog of Spectral Energy Distribution \citep[RCSED][]{RCSED}, we re-fitted a sample of about 800,000 spectra multiple times varying different parameters of the fitting procedure and each step took over a day on a small computer cluster with a single call of the {\sc NBursts} code \citep{CPSA07,CPSK07}, which was derived from {\sc ppxf} and expanded to fit parametric star formation histories and multiple kinematic components. Running even 1000 Monte-Carlo realizations for every spectrum would have turned it into months so that the project would be never finished.

The motivation of this work has two main aspects. On one hand, it is important to have a reliable approach to quickly estimate uncertainties for galaxy kinematics for a large number of spectra coming from spectroscopic surveys. On the other hand, it is often useful to predict the accuracy of kinematics extracted from a spectrum that has been published but has not been made available in a numerical form or has not yet been collected, e.g. while developing strategies for observational campaigns, preparing new observations, writing telescope time proposals, etc. In other words, one needs a simple formula, which can be plugged into a telescope exposure time assuming that a full spectrum fitting technique will be used to extract galaxy kinematics from a spectrum.

We derive the following formulae for uncertainties of $v$ and $\sigma$ obtained in the full spectrum fitting with a purely Gaussian LOSVD $\mathcal{L}(v,\sigma)$
\begin{align}
    \Delta v = \sigma / \sqrt{\sum_{N_{\lambda}} (T * (\mathcal{L}H_1))_i^2 \,\mathrm{SNR}_i^2};\,
    \Delta \sigma = \sigma / \sqrt{\sum_{N_{\lambda}} (T * (\mathcal{L}H_2))_i^2 \,\mathrm{SNR}_i^2} \nonumber \\
    \mathrm{where}\, H_1=x/\sigma;\, H_2=x^2/\sigma^2-1;\, \mathcal{L}(v,\sigma)=\frac{1}{\sigma \sqrt{2\pi}}\exp{\left(-\frac{(x-v)^2}{2\sigma^2}\right)}
    \label{eqerr_gen}
\end{align}
\noindent for a template $T$ spectrum convolved with an instrumental line-spread function of a spectrograph and normalized to its own mean value  ($\mathrm{mean}(T)=1$). Here $*$ denotes a convolution, $\sigma$ is a velocity dispersion value, and the sums are done over all pixels. In a simplified case of constant flux uncertainties corresponding to some mean signal-to-noise ratio in a spectrum, the formulae become:
\begin{align}
    \Delta v = \frac{\sigma}{\mathrm{SNR}} / \sqrt{\sum_{N_{\lambda}}
    (T * (\mathcal{L}H_1))_i^2};\,
    \Delta \sigma = \frac{\sigma}{\mathrm{SNR}} / \sqrt{\sum_{N_{\lambda}} (T * (\mathcal{L}H_2))_i^2}
    \label{eqerr_simple}
\end{align}

\section{Derivation of the Formulae for $v$ and $\sigma$ Uncertainties} \label{sec:derivation}

A pixel space fitting problem using a linear combination of several template spectra can be formulated as a minimization of the following functional \citep{CE04,CPSA07,cappellari17}:
\begin{align}
        \chi^2 = \sum_{N_{\lambda}}\frac{(F_{i}-P_{1p}((T *
        \mathcal{L}(v,\sigma,h_3,h_4))_{i} + P_{2q}) )^2}{\Delta F_{i}^2},\, 
        \mathrm{where}\,
        \quad T_{i} = \sum_{N_{\mathrm{mod}}}k_{i} M_{i}
\label{chi2eq}
\end{align}

\noindent
Here $\mathcal{L}$ is the line-of-sight velocity distribution in the Gauss-Hermite parametrization \citep{1993ApJ...407..525V}; $F_{i}$ and $\Delta F_{i}$ are observed flux and its uncertainty; $T_{i}$ is the flux from a synthetic spectrum, represented by a linear combination of $N_{\mathrm{mod}}$ templates and convolved according to the line-spread function of the spectrograph; $P_{1p}$ and $P_{2q}$ are multiplicative and additive Legendre polynomials of orders $p$ and $q$ for correcting the spectral continuum. For the subsequent calculation we consider that the weights $k_i$ are known and fixed, i.e. we deal with a single template $T_i$. We include multiplicative continuum terms, which match the $T_i$ and $F_i$ flux scales into that template. We consider a purely Gaussian and note that a full Gauss-Hermite LOSVD representation can also be used but the calculations become bulky. We do not include additive continuum terms. Hence, the expression for $\chi^2$ is simplified:
\begin{align}
    \chi^2(v,\sigma) = \sum_{N_{\lambda}}\frac{(F_{i}-(T *
        \mathcal{L}(v,\sigma))_{i})^2}{\Delta F_{i}^2}
\label{chi2eq_simple}
\end{align}

Throughout the calculations we assume that (i) $\chi^2$ changes slowly and monotonically on both sides of the minimum on every parameter; (ii) the parameters are not correlated, which is true for $v$ and $\sigma$; (iii) the number of samples in a spectrum is large enough so that the conversion from discrete to continuum formulation does not change the calculation; (iv) there is no template mismatch, that is a real spectrum is well represented by a template and the flux errors in a real spectrum are Gaussian and correctly estimated so that the normalized by degrees of freedom $\chi^2/D.o.F=1$.

A minimum of $\chi^2$ is reached at the point $(v_0, \sigma_0)$ where partial derivatives turn to zero:
\begin{equation}
    \frac{\partial \chi^2}{\partial v} \bigg\rvert_{v_0}=0;\; \frac{\partial \chi^2}{\partial \sigma}\bigg\rvert_{\sigma_0}=0; 
\label{zerodif}
\end{equation}

To estimate the uncertainties of $v$ and $\sigma$ we use a standard approach from calculus, a Taylor series expansion of $\chi^2$ at $(v_0, \sigma_0)$ to the second degree term and finding a value of a parameter $p$ near the minimum where $\chi^2=\chi^2_{\mathrm{min}}+1$:
\begin{align}
    \chi^2(p_0+\Delta p) = \chi^2(p_0) + 1, \nonumber \\
    \frac{\Delta p^2}{2}\frac{\partial^2 \chi^2}{\partial p^2} \bigg\rvert_{p_0}=1, \nonumber \\
    \Delta p = \sqrt{2}/\sqrt{\frac{\partial^2 \chi^2}{\partial p^2} \bigg\rvert_{p_0}}
\label{eqdeltap}
\end{align}
\noindent
Because here we treat each parameter independently, we use the $\Delta \chi^2=1$ to estimate the uncertainties rather than higher values suggested by the Pearson $\chi^2$ statistics for multiple parameters (e.g. 2.3 for two and 3.5 for three variables, see table~1 in \citealp{1976ApJ...210..642A}).

The second partial derivative of $\chi^2$ from Eq.~\ref{chi2eq_simple} at $(v_0, \sigma_0)$ are expressed as:
\begin{align}
    \frac{\partial^2 \chi^2}{\partial p^2} \bigg\rvert_{p_0}=
    \left\{2\left(\sum \left[(\frac{\partial }{\partial p}(T *
        \mathcal{L})_{i})^2 / \Delta F_i^2\right]\right) -
        2\sum \left[\left(F_{i}-(T *
        \mathcal{L})_{i}\right) \frac{\partial^2}{\partial p^2}(T *
        \mathcal{L})_{i}/\Delta F_i^2\right]\right \}\bigg\rvert_{p_0}
\label{eqd2p}
\end{align}

\noindent
If there is no template mismatch, then at the minimum of $\chi^2$ the second sum becomes zero, because at every pixel in a spectrum the term $(F_{i}-(T *\mathcal{L})_{i})$ is a normally distributed random number with a mathematical expectation of zero. Therefore, we need to compute only first derivatives of a convolution of a template with a LOSVD, and the final result will not depend on the shape of an observed spectrum $F_i$ but rather on a template used to fit it (prior to convolution with a LOSVD) and flux uncertainties $\Delta F_i$. Then the final formula for the uncertainties becomes:
\begin{equation}
    \Delta p = \sqrt{2}/\sqrt{\sum \left[\left(\frac{\partial }{\partial p}(T *
        \mathcal{L})_{i}\right)^2 / \Delta F_i^2\right] \bigg\rvert_{p_0}}
    \label{eqdeltap2}
\end{equation}

Now we can write an integral form of a convolution going from a discrete to continuum approximation ($T_i$ to $T(\lambda)$) and then compute a derivative directly. After trivial calculations, we obtain:
\begin{flalign}
    \frac{\partial }{\partial v}(T(\lambda) *
        \mathcal{L}) = 
        \frac{1}{\sigma^2\sqrt{2\pi}} \int\displaylimits_{-\infty}^{\infty}T(\lambda-x)\frac{(x-v)}{\sigma}\exp{\left (-\frac{(x-v)^2}{2\sigma^2}\right)}dx = \frac{1}{\sigma} \int\displaylimits_{-\infty}^{\infty}T(\lambda-x)\frac{(x-v)}{\sigma}\mathcal{L}(v,\sigma)dx \nonumber \\
    \frac{\partial }{\partial \sigma}(T(\lambda) *
        \mathcal{L}) =         \frac{1}{\sigma^2\sqrt{2\pi}} \int\displaylimits_{-\infty}^{\infty}T(\lambda-x)\left (\frac{(x-v)^2}{\sigma^2}-1\right)\exp{\left (-\frac{(x-v)^2}{2\sigma^2}\right)}dx = \frac{1}{\sigma} \int\displaylimits_{-\infty}^{\infty}T(\lambda-x)\left(\frac{(x-v)^2}{\sigma^2}-1\right)\mathcal{L}(v,\sigma)dx
    \label{eqderconv}
\end{flalign}

We note that each integral in Eq.~\ref{eqderconv} also represents a convolution. In case of $\Delta v$ the kernel is a Gaussian multiplied by the 1-st order Hermite polynomial $H_1=x/\sigma$ and in case of $\Delta \sigma$ it is a Gaussian multiplied by the 2-nd order Hermite polynomial $H_2=(x/\sigma)^2-1$. Now by combining Eq.~\ref{eqderconv} and Eq.~\ref{eqdeltap2} and substituting $\Delta F_i$ with $1/\mathrm{SNR}_i$ under assumptions that the flux scales of $T$ and $F_i$ match and there is no template mismatch, we obtain Eq.~\ref{eqerr_gen} and Eq.~\ref{eqerr_simple}.

\section{Implementation and Code Availability} \label{sec:code}

We implemented the algorithm described above in a form of code in {\sc idl} and {\sc python}. Both implementations yield identical results provided the same input data.

The code implements two different approaches for convolution, a direct pixel-space convolution and a convolution in the Fourier space (default option for the code), which correctly handles undersampled kernels in case of small velocity dispersion values of an order of 1~pix and below \citep[see discussion in][]{cappellari17}. The input parameters for the function {\sc estimate\_pxf\_kin\_err} are: a one-dimensional array of wavelengths, a template spectrum, a value of velocity dispersion, signal-to-noise ratio either per pixel (Eq.~\ref{eqerr_gen}) or a mean value for the whole spectrum (Eq.~\ref{eqerr_simple}) and a keyword to choose an approach for convolution. If a signal-to-noise array (i.e. $F_i/\Delta F_i$) is correctly computed by a data reduction pipeline, then the quality of flux calibration of an observed spectrum will not affect at all the final result.

The {\sc idl} version of the code does not require any third-party dependencies and can be run under {\sc idl} version $>4.0$ or GNU Data Language {\sc gdl} version $>1.0$.

The {\sc python} version of the code works for both {\sc Python 2.7} and {\sc Python 3} and requires only a standard {\sc NumPy} library (version $>1.15.0$). 

\section{Tests Using Monte-Carlo Simulations with Mock Data} \label{sec:MCsim}
\begin{figure}
	\centering
	\includegraphics[width=1\hsize]{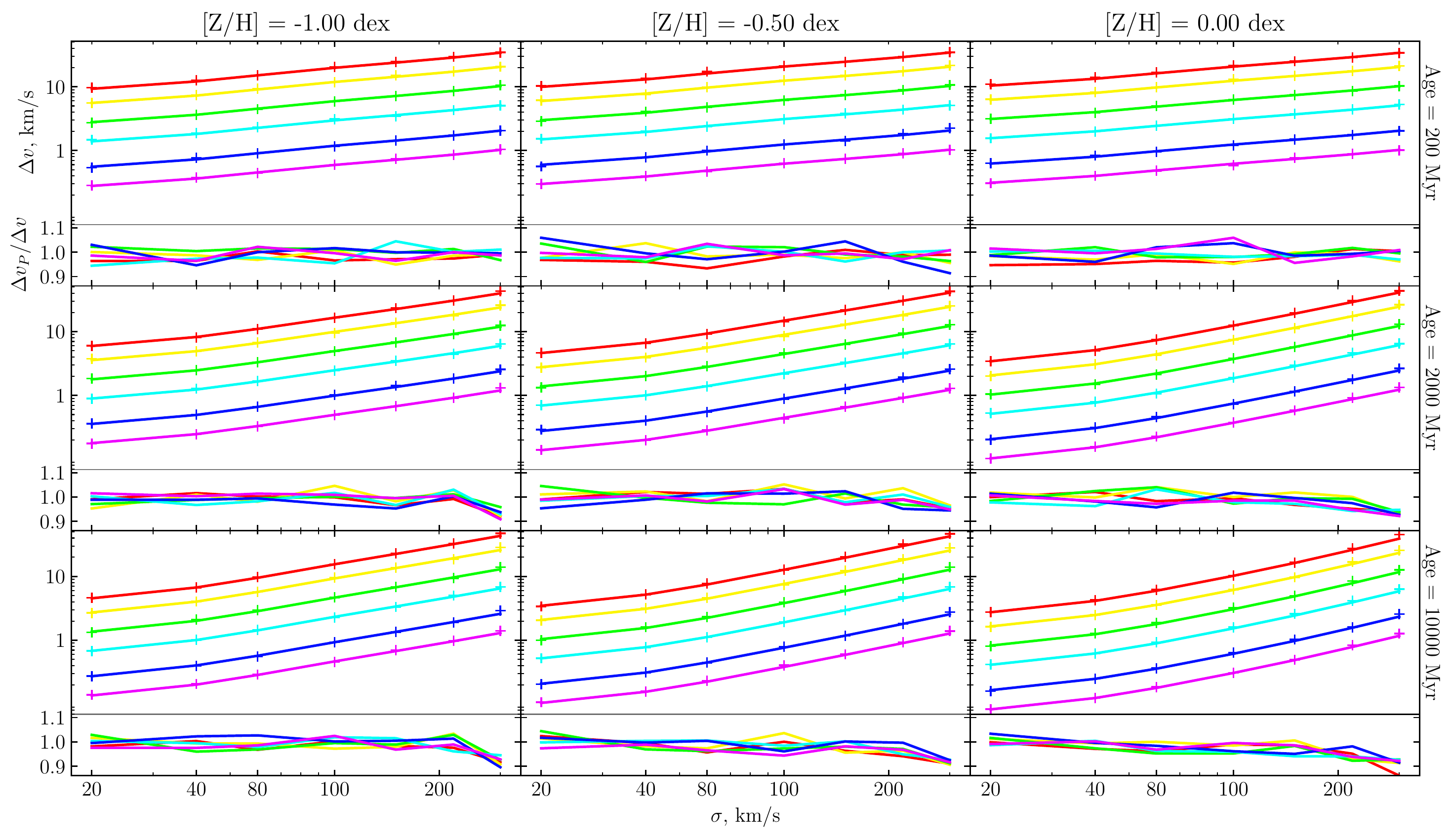}
	\includegraphics[width=1\hsize]{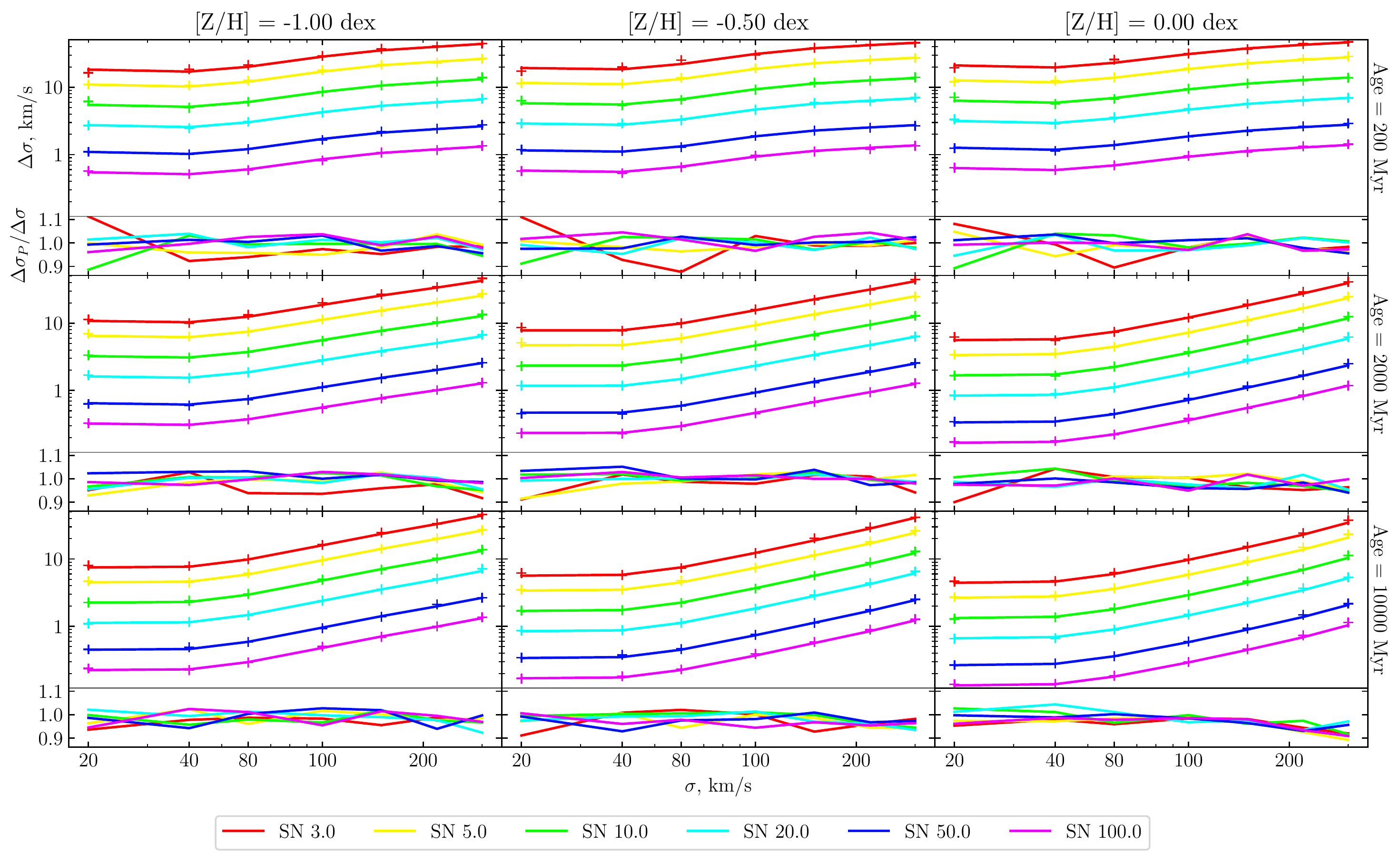}
		\caption{Dependence of the $v$ (top set of panels) and $\sigma$ uncertainties (bottom set of panel) on the velocity dispersion for 9 templates corresponding to different ages and metallicities. Trends for different signal-to-noise ratios are shown using different colors. Plus symbols represent the results of Monte-Carlo simulations with 1000 realizations, and lines are predictions using Eq.~\ref{eqerr_gen}. Narrow panels present a ratio of estimated uncertainties to the results of Monte-Carlo simulations.}
		\label{velsig_MC}
\end{figure}

We ran a suite of Monte-Carlo simulations using mock data to validate Eq.~\ref{eqerr_gen} and demonstrate the consistency between uncertainties predicted by the formulae and ``real'' values computed directly from running a pixel-space fitting code. 

We took 9 simple stellar population templates computed with the {\sc pegase.hr} code \citep{LeBorgne+04} at a spectral resolution $R=10000$ for three different ages, 200~Myr (young), 2000~Myr (intermediate-age), and 10~Gyr (old), and three metallicities [Fe/H]=$-1.0, -0.5, 0.0$~dex. The average depth of absorption lines in galaxy spectra in the optical wavelength range grows when increasing the age and metallicity and the chosen values of stellar population parameters cover a vast majority of cases one has to deal with in real galaxies and star clusters.

We convolved the original templates to a spectral resolution of 25~km~s$^{-1}$, re-binned them to a scale of 20~km~s$^{-1}$ per pixel, and restricted the wavelength range to 3900$<\lambda<$5300~\AA, which roughly correspond to the 1000~gpm intermediate-resolution spectral mode of the Binospec spectrograph \citep{2019PASP..131g5004F}, which we used on several occasions to study internal kinematics and stellar populations of galaxies and star clusters. We chose 7 values of expected velocity dispersion from 20 to 300~km~s$^{-1}$ and 6 values of signal-to-noise ratio between 3 and 100 per pixel.

We ran the {\sc idl} implementation of the {\sc ppxf} code version 5.2.4 (released on 2018/Mar/2) for 1000 noise realizations for every set of parameters (age, metallicity, $\sigma$, SNR), a total of 378,000 simulations, using no additive continuum (\emph{degree=-1}) and the 3rd order multiplicative continuum (\emph{mdegree=3}). We used a corresponding model prior to convolution with the LOSVD as a template to avoid template mismatch by construction. We defined a noise vector for every spectrum as a set of normally distributed random numbers with the dispersion $F_i/\mathrm{SNR}$ in the $i$-th pixel. Then we computed uncertainties of $v$ and $\sigma$ for every numerical experiment as a standard deviation of $v$ and $\sigma$ reported by {\sc ppxf} from 1000 realizations. In the end, we compare the uncertainties predicted by Eq.~\ref{eqerr_gen} to the results of simulations. 

The results of simulations are presented in Fig.~\ref{velsig_MC}. The top set of panels shows estimated uncertainties of $v$ and the bottom panels are for $\sigma$. For every age and metallicity there are two panels, which show values of uncertainties (top) and ratios between predicted and computed uncertainties (bottom). The colored lines correspond do different values of the input signal-to-noise ratio. 

We see excellent agreement within a few per~cent between the uncertainties predicted by our approach and derived from Monte-Carlo simulations. One should keep in mind that 1000 realizations yield an internal statistical accuracy of uncertainties of 1.6~\%. In a few cases especially for $\Delta v$ we see a slight trend towards high velocity dispersion where Eq.~\ref{eqerr_gen} seem to under-predict the uncertainties by up-to 10~per~cent. This, however, might be the result of a worse convergence of the non-linear minimization used in {\sc ppxf}, which lead to a higher spread of solutions. We attribute a larger disagreement for low-SNR simulations (3 and 5) to the same effect.

\section{Discussion}
\subsection{Dependence of uncertainties on a shape of a spectrum}

\begin{figure}
    \centering
    \includegraphics[width=0.45\hsize]{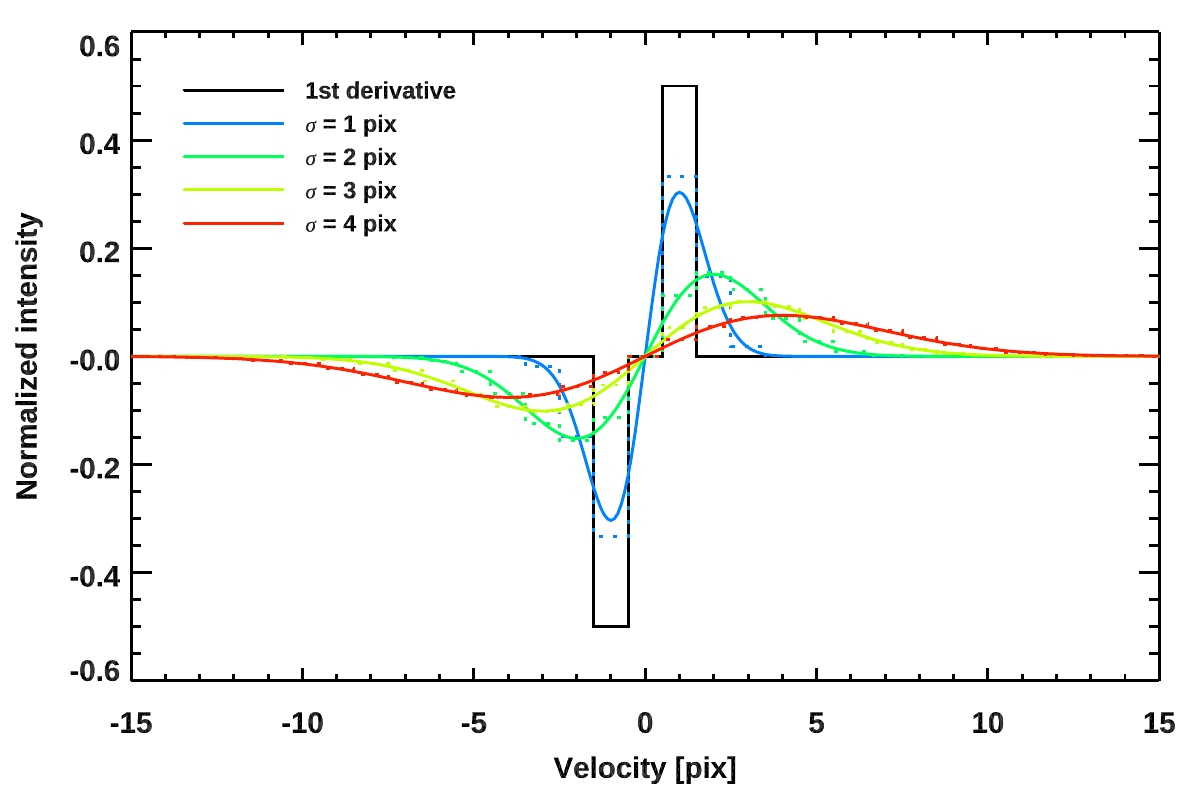}
    \includegraphics[width=0.45\hsize]{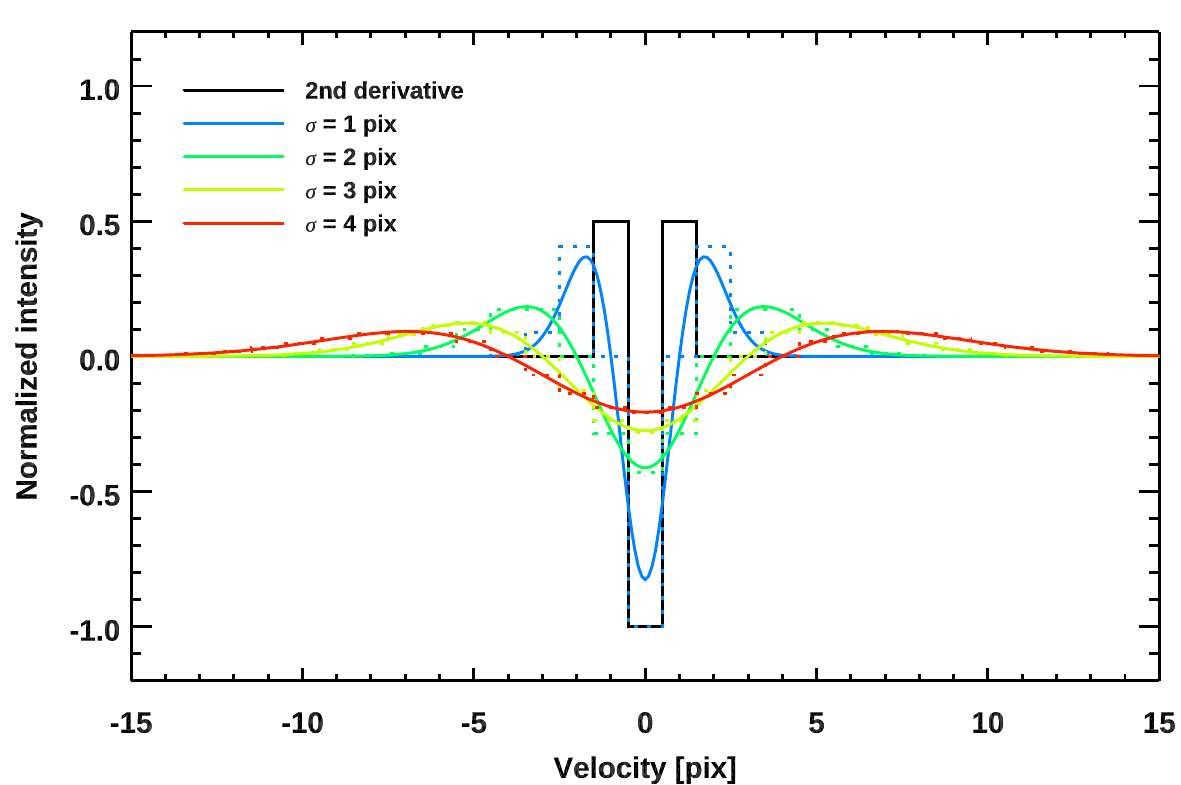}
    \caption{Convolution kernels for the calculation of uncertainties of radial velocity (left) and velocity dispersion (right). Black solid histograms show convolution kernels for the computation of the 1st (left) and 2nd (right) derivatives using finite differences. Colored lines show the convolution kernels $H_1 \mathcal{L}$ (left) and $H_2 \mathcal{L}$ (right) for several values of the velocity dispersion (see insets for details). Colored dotted histograms show the discrete representation of the corresponding convolution kernels.}
    \label{fig:krnl}
\end{figure}

The computation of uncertainties presented in Eq.~\ref{eqerr_gen} has some interesting properties. 
\begin{itemize}
    \item Uncertainties on both parameters increase when $\sigma$ grows, which is easy to explain: higher velocity dispersion smooths the spectrum making both $v$ and $\sigma$ more uncertain.
    \item The convolution kernels $\mathcal{L}H_1$ and $\mathcal{L}H_2$ both have zero total values, therefore a sum of all pixels in a template spectrum convolved with them will be close to zero. However, because the convolution result is squared, the sum will always be positive.
    \item The convolution $(T_i * (\mathcal{L}H_1))$ is in fact a two-sided first numerical derivative of a template $T_i$ in terms of finite differences convolved with a Gaussian $\mathcal{L}$ (see Fig.~\ref{fig:krnl} left). The convolution and numerical derivation can be done in any order because it is equivalent to a double convolution with the $(-0.5,0,+0.5)$ kernel and a Gaussian; and a convolution is commutative. Therefore, a local gradient of a convolved template spectrum determines the quality of radial velocity determination: deep broad and narrow spectral lines and sharp continuum breaks both improve the quality.
    \item The convolution $(T_i * (\mathcal{L}H_2))$ is a second  derivative of $T_i$ in terms of finite differences expressed as a convolution of a template with the $(+0.5,-1,+0.5)$ kernel then convolved with a Gaussian $\mathcal{L}$ (see Fig.~\ref{fig:krnl} right). Therefore, a local curvature of a convolved template spectrum determines the quality of velocity dispersion measurements: deep narrow spectral lines are crucial, while broad lines and continuum breaks do not improve the quality.
\end{itemize}

One can see a few notable trends in Fig.~\ref{velsig_MC}. 
\begin{itemize}
    \item In the selected wavelength range, which is quite common for extragalactic observations, the uncertainties of $v$ and $\sigma$ are similar in absolute values when $\sigma > \sigma_{\mathrm{LSF}}$. 
    \item The instrumental resolution (25~km~s$^{-1}$) puts an effective lower limit on the measurement of $\sigma$ (see the curves flatten at lower $\sigma$ values below the instrumental resolution). This is trivially explained when one keeps in mind that the instrumental resolution is a convolution of a ``true'' template spectrum with a line-spread-function (Gaussian in our case). Using the commutativity of the convolution and the property of a Gaussian dispersion during convolution ($\sigma_{\mathrm{conv}}^2 = \sigma_{\mathrm{orig}}^2 + \sigma_{\mathrm{kernel}}^2$), we conclude that this behavior of uncertainties is an effect of simple error propagation, because the spectrum fitting effectively measures $\sigma_{\mathrm{galaxy}}^2 + \sigma_{\mathrm{LSF}}^2$.
    \item However, the flattening is much less pronounced for radial velocities, which one would expect from the fact that $\Delta v$ depends on a local gradient of the spectral shape, which is much less affected by smoothing than a local curvature, that $\Delta \sigma$ depends on. 
    \item The absolute values of uncertainties grow towards younger ages and lower metallicities, which illustrates the fact that absorption features become less pronounced.
    \item The uncertainties are inversely proportional to the signal-to-noise ratio even in very noisy spectra at SNR$=3-5$.
\end{itemize}

\subsection{Real Data Examples and Potential Caveats}

We started the development of our approach while analyzing low signal-to-noise spectral data collected with Binospec at the 6.5-m converted Multiple Mirror Telescope for low surface brightness galaxies in the Coma cluster published in \citet{Chilingarian+19}. Our velocity dispersion estimates from intermediate-resolution ($R\approx4,800$) spectra having low signal-to-noise ratio of 4--5 per pixel were questioned by colleagues, in particular by comparing our $\sigma$ uncertainties (7--9~km~s$^{-1}$) to those (7~km~s$^{-1}$) obtained by \citet{2016ApJ...828L...6V} for a similar ultra-diffuse galaxy DF~44 using a larger 10-m Keck telescope and much longer integration time (33.5~h vs 2~h) that yielded a substantially higher signal-to-noise ratio of 14 per pixel. The key difference was a spectral configuration and a wavelength range used in the two studies. We used a blue optical setup covering a lot of prominent absorption lines in a wide wavelength range (3900$<\lambda<$5300~\AA, same as used in the Monte-Carlo simulations presented earlier), while \citet{2016ApJ...828L...6V} used a narrow spectral region centered on H$\alpha$ (6444$<\lambda<$6679~\AA), which contains very few absorption lines. We applied Eq.~\ref{eqerr_simple} to a stellar population template having age of 10~Gyr and [Fe/H]=$-1.5$~dex representative of a spectrum of DF~44 with all the parameters reported in the paper ($\sigma$, spectral resolution, sampling, etc.) and obtained an estimated $\sigma$ uncertainty of 6.5~km~s$^{-1}$ fully consistent with 7~km~s$^{-1}$ presented by \citet{2016ApJ...828L...6V}. 

Similarly, we applied the formulae to the Binospec setup used by \citet{Chilingarian+19} using signal-to-noise ratios and stellar population parameters reported in the paper and obtained the uncertainty estimates for both $v$ and $\sigma$ fully consistent with the published values, which were taken directly from the minimization routine. Moreover, we tried both Eq.~\ref{eqerr_simple} for average values of signal-to-noise and Eq.~\ref{eqerr_gen} with the flux uncertainties provided by the Binospec data reduction pipeline \citep{2019PASP..131g5005K}. The results were generally consistent but the use of Eq.~\ref{eqerr_gen} provided better agreement with the values from the minimization routine.

The derived expressions for uncertainties of kinematics quantify the amount of $v$ and $\sigma$-sensitive information in an absorption-line spectrum, which can potentially be extracted from it. One can obtain this quantity empirically by computing derivatives $\partial \chi^2/\partial v$ and $\partial \chi^2/\partial \sigma$ from a template grid using finite differences. A similar approach was used by \citet{Chilingarian09,CMHI11} to assess the quantity of age- and metallicity-sensitive information per wavelength bin. This information can be used to choose the optimal setup for a spectrograph to measure velocity dispersion or to choose the best spectral range for the data analysis. For example, \citet{2013PASP..125.1362F} empirically determined the optimal wavelength range to extract velocity dispersions from low-resolution galaxy spectra by running the full spectrum fitting code many times and adjusting the wavelength range. Using Eq.~\ref{eqerr_gen} one can do it without running the code by simply analyzing template spectra.

There are several caveats of our approach if one plans to get quantitatively correct estimates of uncertainties as a quick alternative to Monte-Carlo simulations using Eq.~\ref{eqerr_gen}: (i) flux errors have to be correctly estimated and propagated through a data reduction pipeline that is used to produce spectra; (ii) strong template mismatch would affect the estimates of uncertainties, they will get underestimated; (iii) there is a degeneracy between $\sigma$ and [Fe/H] \citep[see Appendix A in][]{CPSA07} when using stellar population models, which lead to increased $\sigma$ uncertainties, e.g. their underestimation by Eq.~\ref{eqerr_gen}; (iv) there is a degeneracy between $v$ and Gauss-Hermite $h3$ and as well between $\sigma$ and $h4$, which would also lead to the underestimation of uncertainties by Eq.~\ref{eqerr_gen} if one uses a 4-th order LOSVD expansion in the analysis.

Despite all the caveats, the presented solution estimates $v$ and $\sigma$ uncertainties quickly and precisely and it can be used in a large range of situations from validating published kinematics of galaxies to preparing observational programs.

\acknowledgments

We thank D.~Fabricant, I.~Katkov, and A.~Afanasiev for fruitful discussions. IC is supported by the Telescope Data Center at Smithsonian Astrophysical Observatory. The authors acknowledge the Russian Science Foundation grant 19-12-00281 and the Program of development of M.V. Lomonosov Moscow State University for the Leading Scientific School ``Physics of stars, relativistic objects and galaxies''.

\bibliography{kin_err_pasp}{}
\bibliographystyle{aasjournal}

\end{document}